\documentclass[epj]{svjour}

\usepackage{graphicx}
\usepackage{fancyhdr}

\def\makeheadbox{{%  remove EPJC header
 }}

\def\mytitle{My title}
\def\myauthors{My name}
\def\mytype{My type of session}
\def\mysession{My session}
\def\mytitle{QCD resummations for slepton pair hadroproduction} %Put your title here!
\def\myauthors{Benjamin Fuks}    %Put your name here!
\def\mytype{Contributed Talk}
\def\mysession{Colliders - SUSY Phenomenology}
\def\d{{\rm d}}
\def\bea{\begin{eqnarray}}
\def\eea{\end{eqnarray}}
\def\as{\alpha_s}
\def\nn{\nonumber}

\begin{document}
\title{Transverse-momentum, threshold and joint resummations for
slepton pair production at hadron colliders}
% \subtitle{Do you have a subtitle?\\ If so, write it here}
\author{Benjamin Fuks
\thanks{\emph{Email:} benjamin.fuks@physik.uni-freiburg.de}%
}                     % Do not remove

\institute{Laboratoire de Physique Subatomique et de Cosmologie,
Universit\'e Joseph Fourier/CNRS-IN2P3,\\ \vspace{.1cm}53 Avenue
des Martyrs, F-38026 Grenoble, France\\ Physikalisches Institut,
Albert-Ludwigs-Universit\"at Freiburg, \\ Hermann-Herder-Stra\ss e
3, D-79104 Freiburg i.Br., Germany}

% Preprint number:  FREIBURG PHENO-07-06, LPSC 07-99

\date{}
\abstract{We present precision calculations of the
transverse-momentum spectrum and the invariant-mass distribution
for slepton pair production at hadron colliders. We implement the
transverse-momentum, threshold and joint resummation formalisms at
the next-to-leading logarithmic accuracy and consistently match
the obtained result with the pure perturbative result at the first
order in the strong coupling constant, i.e.\ at
$\mathcal{O}(\as)$. We give numerical predictions for selectron
and stau pair production, and compare the various resummed cross
sections with the perturbative result.
\PACS{
      {12.60.Jv}{Supersymmetric models} \and
      {14.80.Ly}{Supersymmetric partners of known particles} \and
      {12.38.Cy}{Summation of perturbation theory}
     } % end of PACS codes
} %end of abstract
\maketitle
%
%%%%%%%%%%%%%%%%%%%%%%%%%%%%%%%%%%%%%%%%%%%%%%%%%%%%%%%%%%%%%%%%

\section{Introduction}
\label{intro}

One of the main goals of the experimental programme at present and
future hadron colliders is to perform an extensive and conclusive
search of the supersymmetric (SUSY) partners of the Standard Model
(SM) particles predicted by the Minimal Supersymmetric Standard
Model \cite{Nilles:1983ge,Haber:1984rc}. Scalar leptons are among
the lightest supersymmetric particles in many SUSY-breaking
scenarios, and often decay into the corresponding SM partner and
the lightest stable SUSY particle, the distinctive signature at
hadron colliders consisting thus in a highly energetic lepton pair
and associated missing energy. Corresponding production cross
sections have been extensively studied at leading order (LO)
\cite{Dawson:1983fw,Chiappetta:1985ku,delAguila:1990yw,Baer:1993ew,%
Bozzi:2004qq} and also at next-to-leading order (NLO)
\cite{Baer:1997nh,Beenakker:1999xh,Bozzi:2007qr} of perturbative
QCD.

The aim of this work is to perform an accurate calculation of the
transverse-momentum ($q_T$) spectrum and to investigate the
threshold-enhanced contributions, including soft-gluon resummation
at the next-to-leading logarithmic (NLL) accuracy
\cite{Bozzi:2007qr,Bozzi:2006fw,Bozzi:2007te}. This allows for
the reconstruction of the mass and the determination of the spin
of the produced particles by means of the Cambridge (s)transverse
mass variable \cite{Lester:1999tx,Barr:2005dz} and for
distinguishing thus the SUSY signal from the SM background, mainly
due to $WW$ and $t \bar t$ production \cite{Lytken:2003ed,%
Andreev:2004qq}.

When studying the transverse-momentum distribution of a produced
slepton pair with an invariant mass $M$, it is convenient to
separate the large- and small-$q_T$ regions. For the large values
of $q_T$ the use of the fixed-order perturbation theory is fully
justified, since the perturbative series is controlled by a small
expansion parameter, $\alpha_s(M^2)$, but in the small-$q_T$
region, where the bulk of the events will be produced, the
coefficients of the perturbative expansion are enhanced by powers
of large logarithmic terms, $\ln(M^2 / q_T^2)$. As a consequence,
results based on fixed-order calculations diverge as $q_T \to 0$,
and the convergence of the perturbative series is spoiled.
Furthermore, at the production threshold, when the initial partons
have just enough energy to produce the slepton pair in the final
state, the mismatch between virtual corrections and phase-space
suppressed real-gluon emission leads also to the appearance of
large logarithmic terms $[\ln(1-z)/(1-z)]_+$ where $z=M^2/s$, $s$
being the partonic centre-of-mass energy. However, the convolution
of the partonic cross section with the steeply falling parton
distributions enhances the threshold contributions even if the
hadronic threshold is far from being reached, i.e.\ $\tau=M^2/S
\ll 1$, where $S$ is the hadronic centre-of-mass energy, and in
this intermediate $z$ region, large corrections are expected
for the Drell-Yan production of a rather light slepton pair at the
CERN Large Hadron Collider (LHC). Accurate calculations of
$q_T$-spectrum and invariant mass distribution must then include
soft-gluon resummation in order to obtain reliable perturbative
predictions and properly take these logarithmic terms into
account.

The methods to perform all-order soft-gluon resummation at small
$q_T$ \cite{Catani:2000vq,Bozzi:2005wk} and at large $z$
\cite{Sterman:1986aj,Catani:1989ne,Kramer:1996iq,Catani:2001ic}
are well known. However, since the dynamical origin of the
enhanced contributions is the same both in transverse-momentum and
in threshold resummations, a joint resummation formalism has been
developed in the last eight years, resumming the $\ln(M^2 /
q_T^2)$ and $[\ln(1-z)/(1-z)]_+$ terms simultaneously
\cite{Bozzi:2007te,Laenen:2000ij,Kulesza:2002rh}.

%%%%%%%%%%%%%%%%%%%%%%%%%%%%%%%%%%%%%%%%%%%%%%%%%%%%%%%%%%%%%%%%

\section{Resummation formalisms at the next-to-leading logarithmic
level} \label{sec:2}

In Mellin $N$-space, the hadronic cross section for the hard
scattering process \bea h_a(p_a) \, h_b(p_b) \to \tilde{l}
\tilde{l}^\ast(M, q_T) + X,~\eea where a slepton pair with an
invariant mass $M$ and a transverse momentum $q_T$ is produced,
naturally factorizes \bea \frac{\d^2\sigma}{\d M^2\, \d q^2_T} &=&
\sum_{a,b} \oint_{\cal C} \frac{\d N}{2 \pi i}\, \tau^{-N}\,
f_{a/h_a}(N\!+\!1)\, f_{b/h_b}(N\!+\!1)\nn\\&& \times
\frac{\d^2\hat\sigma_{ab}}{\d M^2 \d q_T^2}(N).~\eea $f_{a,b/
h_{a,b}}$ are the $N$-moments of the universal distribution
functions of partons $a,b$ inside the hadrons $h_{a,b}$,
$\hat{\sigma}_{ab}$ the relevant partonic cross section, and the
contour ${\cal C}$ in the complex $N$-space will be specified in
Sec.\ \ref{sec:3}. The dependence on the unphysical
renormalization and factorization scales $\mu_R$ and $\mu_F$ has
been removed for brevity. In general, the corresponding partonic
cross section can be written as \bea \d\sigma =
\d\hat{\sigma}^{(\rm res.)} + \d \hat{\sigma}^{(\rm fin.)},~\eea
where the resummed contribution is given by \bea \frac{\d{\hat
\sigma}_{ab}^{(\rm res.)}}{d M^{2} d q_T^2}(N) &=& \int
\frac{b}{2}\, \d b\, J_0(b\,q_T)\, \mathcal{W}_{ab}(N, b),~\\
\frac{\d\hat\sigma_{ab}^{(\rm res.)}}{\d M^2}(N) &=&
{\hat\sigma}^{{\rm (res)}}_{ab}(N),~ \eea for $q_T$ spectrum and
invariant-mass distribution, respectively. The impact-parameter
$b$ is the variable conjugate to $q_T$ through a Fourier
transformation, and $J_0(x)$ is the $0^{{\rm th}}$-order Bessel
function. The perturbative functions ${\cal W}$ and ${\hat\sigma}$
embody the all-order dependence on the large logarithms and can be
expressed in an exponential form, \bea {\cal
W}_{ab}(N,b) &=& {\cal H}_{ab}(N) \exp\{{\cal G}(N,b)\},~\\
\hat\sigma^{({\rm res})}_{a b}(N) &=& \sigma^{(LO)}\, \tilde{C}_{a
b}(N)\, \exp\{{\cal G}(N)\}, \eea where $\sigma^{(LO)}$ is the LO
cross section. The process-in\-de\-pen\-dent Sudakov form factor
${\cal G}$ allows to resum the soft-collinear radiation, while the
process-dependent functions ${\cal H}$ and $\tilde{C}$ contain all
the terms due to hard virtual corrections and collinear radiation.
The general expressions of these functions can be found in Ref.\
\cite{Bozzi:2006fw} for transverse-momentum resummation, in Ref.\
\cite{Bozzi:2007qr} for threshold resummation and in Ref.\
\cite{Bozzi:2007te} for joint resummation, so that an analysis at
the NLL accuracy can be performed.

%%%%%%%%%%%%%%%%%%%%%%%%%%%%%%%%%%%%%%%%%%%%%%%%%%%%%%%%%%%%%%%%

\section{Inverse transform and matching} \label{sec:3}

Once resummation has been achieved in $N$- and $b$-space (if
relevant), inverse transforms have to be performed in order to get
back to the physical spaces. Special attention has to be paid to
the singularities in the resummed exponent, and the integration
contours of the inverse transforms must therefore avoid hitting
any of these poles. The $b$-integration is performed by deforming
the integration contour with a diversion into the complex
$b$-space \cite{Laenen:2000de}, while the inverse Mellin transform
is performed following a contour inspired by the Minimal
Prescription \cite{Catani:1996yz} and the Principal Value
Resummation \cite{Contopanagos:1993yq}.

In order to keep the full information contained in the fixed-order
calculation and to avoid possible double-counting of the
logarithmically enhanced contributions, a matching procedure of
the NLL resummed cross section to the $\mathcal{O}(\as)$ result is
performed through the formulae \bea && \frac{\d^2\sigma}{\d
M^2\,\d q_T^2} = \frac{\d^2\sigma^{({\rm F.O.})}}{\d M^2\,\d
q_T^2} + \oint_{C_N} \frac{\d N}{2\pi i}\, \tau^{-N}\!\! \int
\frac{b\, \d b}{2} J_0(b\, q_T) \nn\\&&\,\, \,\, \times
\left[\frac{\d^2\sigma^{{\rm (res)}}}{\d M^2\,\d q_T^2}(N, b; \as)
- \frac{{\rm d}^2\sigma^{{\rm (exp)}}}{\d M^2\,\d q_T^2}(N, b;
\as) \right],~\\
&&\frac{\d\sigma}{\d M^2}(\tau, M) = \frac{\d\sigma^{({\rm
F.O.})}}{\d M^2}(\tau, M)  + \oint_{C_N} \frac{\d N}{2\pi i}\,
\tau^{-N}\nn\\&&\,\,\,\, \times \left[\frac{\d\sigma^{{\rm
(res)}}}{\d M^2}(N, M) - \frac{{\rm d}\sigma^{{\rm (exp)}}}{\d
M^2}(N, M) \right] \eea for $q_T$ spectrum and invariant-mass
distribution, respectively. $\d^2\sigma^{({\rm F.O.})}$ is the
fixed-order perturbative result, $\d^2\sigma^ {({\rm res})}$ is
the resummed cross section and $\d^2\sigma^{({\rm exp})}$ is the
truncation of the resummed cross section to the same perturbative
order as $\d^2\sigma^{({\rm F.O.})}$.

%%%%%%%%%%%%%%%%%%%%%%%%%%%%%%%%%%%%%%%%%%%%%%%%%%%%%%%%%%%%%%%%

\section{Numerical results} \label{sec:4}

\begin{figure}
\includegraphics[width=0.45\textwidth]{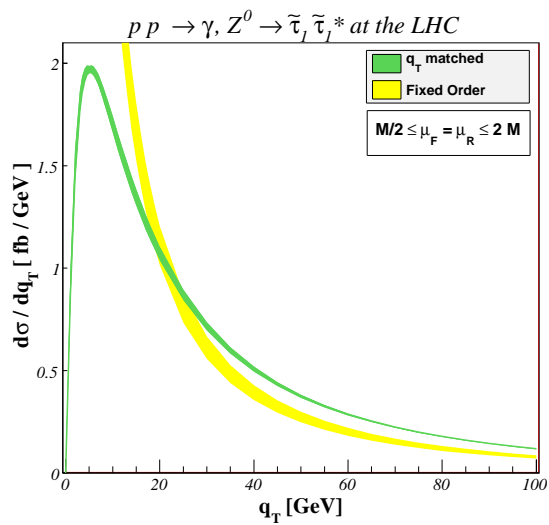}
\caption{Differential cross section for the process $p p \to
\tilde{\tau}_{1}\tilde{\tau}_{1}^{*}$ at the LHC for the benchmark
scenario SPS 7. NLL+LO matched result (with $q_T$ resummation) and
$\mathcal{O}(\as)$ result are shown.}
\label{fig:1}       % Give a unique label
\end{figure}

We now present numerical results for slepton pair production at
the LHC, with an hadronic centre-of-mass energy of $\sqrt{S} = 14$
TeV. For the LO (NLO and NLL) predictions, we use the LO 2001
\cite{Martin:2002dr} (NLO 2004 \cite{Martin:2004ir}) MRST sets of
parton distribution functions and $\alpha_s$ is evaluated at two-loop
accuracy. In the following, we choose the typical SUSY benchmark
points SPS 1a, SPS 7 \cite{Allanach:2002nj} and BFHK B
\cite{Bozzi:2007me} which gives, after the renormalization group
evolution of the SUSY-breaking parameters performed by the {\tt
SPheno} \cite{Porod:2003um} or {\tt SuSpect} \cite{Djouadi:2002ze}
programmes, light sleptons of 100-200 GeV and rather heavy squarks
with masses in the 500-1000 GeV range.

In Fig.\ \ref{fig:1}, we show the $q_T$-spectrum of a
$\tilde{\tau}_1$ pair, for the benchmark point SPS 7, and we allow
$\mu_{F}=\mu_{R}$ to vary between $M/2$ and $2M$ to estimate the
perturbative uncertainty. We also integrate the equations of the
previous sections with respect to $M^2$, taking as lower limit the
energy threshold for $\tilde{\tau}_1 \tilde{\tau}_1^\ast$
production and as upper limit the hadronic energy. The
$\mathcal{O}(\as)$ result diverges, as expected, as $q_{T}\to 0$,
while the effect of resummation is clearly visible for small and
intermediate values of $q_T$, the resummation-improved result
being nearly 39\% higher at $q_T=50$ GeV than the pure fixed order
result. Let us note that when integrated over $q_T$, the former
leads to a total cross section of 66.8 fb in good agreement with
the QCD-corrected total cross section at ${\cal O}(\alpha_s)$. The
scale dependence is clearly improved with respect to the pure
fixed-order calculations, being about 10\% for the fixed order
result, while it is always less than 5\% for the matched curve.

\begin{figure}
\includegraphics[width=0.45\textwidth]{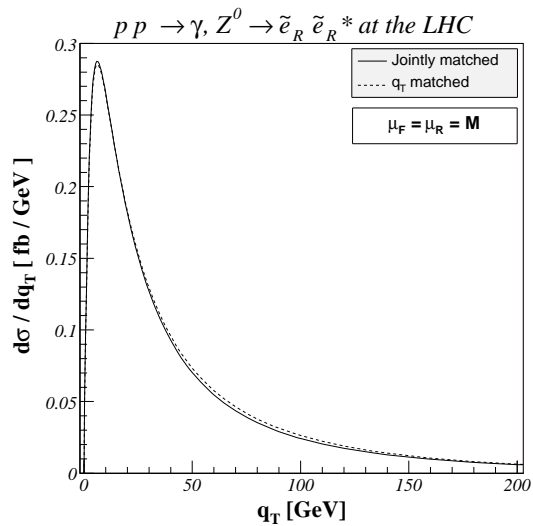}
\caption{Transverse-momentum distribution for selectron pair
production at the LHC in the framework of joint (full) and $q_T$
(dotted) resummations, for the benchmark point BFHK B.}
\label{fig:2}       % Give a unique label
\end{figure}

In Fig.\ \ref{fig:2}, we compare the jointly- and $q_T$-matched
results for the production of a right-handed selectron pair, for
the benchmark point BFHK B. The behaviour of the two curves is
similar in the small-$q_T$ region, while the jointly-resummed
cross section is about 5\%-10\% lower than the $q_T$-resum\-med
one for intermediate values of the transverse momentum $q_T$. This
effect is thus clearly related to the threshold-enhanced
contributions important in the large-$M$ region, which are not
present in $q_T$ resummation.

\begin{figure}
\includegraphics[width=0.45\textwidth]{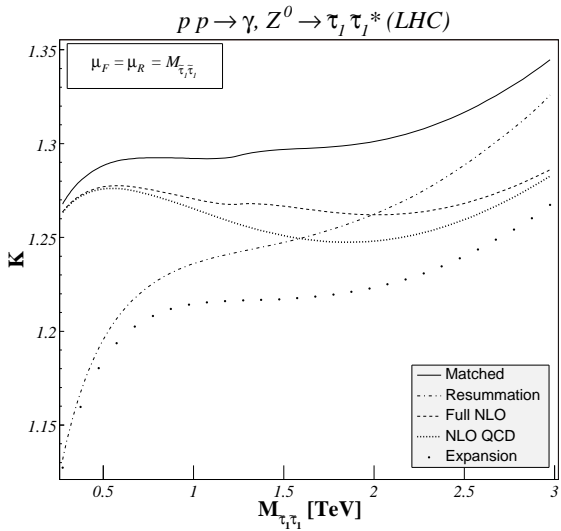}
\caption{$K$-factors as defined in Eq.\ (\ref{eq:K}) for
$\tilde{\tau}_1$ pair production at the LHC for the benchmark
point SPS 1a. We show the total NLL+NLO (threshold) matched
result, the (threshold) resummed result at NLL, the fixed order
NLO SUSY-QCD and QCD results, and the resummed result expanded up
to NLO.}
\label{fig:3}       % Give a unique label
\end{figure}

For the scenario SPS 1a and $\tilde{\tau}_1$ pair production, we
show in Fig.\ \ref{fig:3} the cross section correction factors
\bea \label{eq:K} K^i = \frac{{\rm d}\sigma^i / {\rm d}M}{{\rm
d}\sigma^{\rm LO} / {\rm d}M}, \eea where $i$ labels the
corrections induced by NLO QCD, additional NLO SUSY-QCD,
resummation, the mat\-ched contributions as well as the
fixed-order expansion of the resummation contribution as a
function of the invariant mass $M$. In the left part of this plot,
the matched result is less than 0.5\% larger than the NLO
(SUSY-)QCD result, since the slepton pair is produced with a
relatively small invariant mass compared to the total available
centre-of-mass energy, so that $z\ll 1$ and the resummation of
($1-z$)-logarithms is less important. Finite terms dominate the
cross section and the resummed contribution is close to its
fixed-order expansion. Only at large $M$ the logarithms become
important and lead to a 7\% increase of the $K$-factor with
resummation over the fixed-order result. In this region, the cross
section is dominated by the large logarithms, and the resummed
result approaches the total prediction, while the NLO QCD
calculation approaches the expanded resummed result. In the
intermediate-$M$ region, we are still far from the hadronic
threshold region and both resummed and fixed-order contributions
are needed, a consistent matching being thus mandatory.

\begin{figure}
\includegraphics[width=0.45\textwidth]{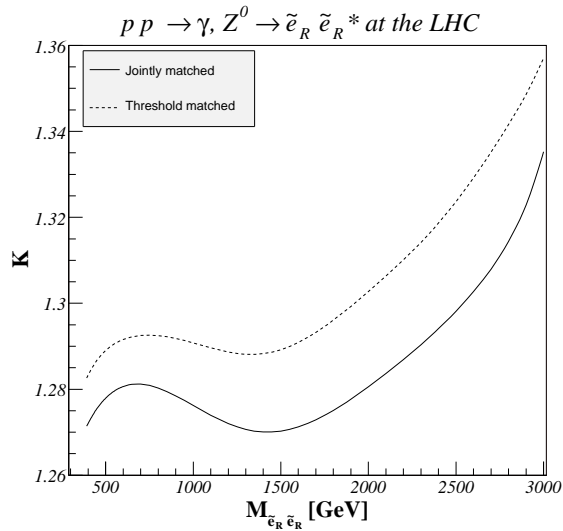}
\caption{$K$-factors as defined in Eq.\ (\ref{eq:K}) for
$\tilde{e}_R$ pair production at the LHC for the benchmark point
BFHK B. We show the total NLL+NLO jointly (full), and threshold
(dashed) matched results.}
\label{fig:4}       % Give a unique label
\end{figure}

In Fig.\ \ref{fig:4}, we eventually show the differences between
threshold and joint resummations for ${\tilde e}_R$ pair
production within the benchmark scenario BFHK B, which are only
about one or two percents. These are due to the choices of the
Sudakov form factor and of the $\mathcal H$-function, which
correctly reproduce transverse-momentum resummation in the limit
of $b\to\infty$, $N$ being fixed, but which present some
differences with pure threshold resummation in the limit $b\to 0$
and $N\to\infty$, as it was the case for joint resummation for
Higgs and electroweak boson production
\cite{Kulesza:2002rh,Kulesza:2003wn}. However, this effect is
under good control, since it is much smaller than the theoretical
scale uncertainty of about 7\%.

%%%%%%%%%%%%%%%%%%%%%%%%%%%%%%%%%%%%%%%%%%%%%%%%%%%%%%%%%%%%%%%%

\section{Conclusions}\label{sec:5}
Within this work, soft-gluon resummation effects are now
consistently included in predictions for various distributions for
slepton pair production at hadron colliders, exploiting the $q_T$,
threshold, and joint resummation formalisms. We found that the
effects obtained from resumming the enhanced soft contributions
are important, even far from the critical kinematical regions
where the resummation procedure is fully justified. The numerical
results show a considerable reduction of the scale uncertainty
with respect to fixed order results, these features leading then
to an increased stability of the perturbative results and thus to
a possible improvement of the slepton pair search strategies at
the LHC.

%%%%%%%%%%%%%%%%%%%%%%%%%%%%%%%%%%%%%%%%%%%%%%%%%%%%%%%%%%%%%%%%

\end{document}